\documentclass[twocolumn]{aastex6}

\usepackage{verbatim}
\usepackage{aas_macros}
\usepackage{hyperref}
\usepackage{natbib}
\usepackage{graphicx}
\usepackage{epsfig}
\usepackage{amsmath}
\usepackage[english]{babel}

\begin{document}


\title{Supermassive black holes as the regulators of star formation in central galaxies}

\author{Bryan A. Terrazas\altaffilmark{1}, Eric F. Bell\altaffilmark{1}, Joanna Woo\altaffilmark{2}, Bruno M. B. Henriques\altaffilmark{2}$^{,}$\altaffilmark{3}}
\shortauthors{Terrazas et al.}

\altaffiltext{1} {Department of Astronomy, University of Michigan, Ann Arbor, MI 48109, USA}
\altaffiltext{2} {Department of Physics, Institute for Astronomy, ETH Zurich, 8093 Zurich, Switzerland}
\altaffiltext{3} {Max-Planck-Institut f{\"u}r Astrophysik, D-85741 Garching, Germany}

\begin{abstract}
We present a relationship between the black hole mass, stellar mass, and star formation rate of a diverse group of 91 galaxies with dynamically-measured black hole masses. For our sample of galaxies with a variety of morphologies and other galactic properties, we find that the specific star formation rate is a smoothly decreasing function of the ratio between black hole mass and stellar mass, or what we call the specific black hole mass. In order to explain this relation, we propose a physical framework where the gradual suppression of a galaxy's star formation activity results from the adjustment to an increase in specific black hole mass and, accordingly, an increase in the amount of heating. From this framework, it follows that at least some galaxies with intermediate specific black hole masses are in a steady state of partial quiescence with intermediate specific star formation rates, implying that both transitioning and steady-state galaxies live within this region known as the ``green valley." With respect to galaxy formation models, our results present an important diagnostic with which to test various prescriptions of black hole feedback and its effects on star formation activity.
\end{abstract}




\section{Introduction}
\label{sec:Intro}

Large scale galaxy surveys have made it clear that there has been a pronounced growth in the number of galaxies that host little to no star formation \citep{bwm2004, bdj2007, fww2007, mms2013, iml2013, tqt2014, mch2015}, reflecting the overall declining cosmic star formation rate observed in the universe since $z = 2$ (see \citealt{md2014} for a review). In an effort to understand how the quiescent population grows, these observational studies have used color-magnitude diagrams to split galaxies into a blue cloud and red sequence. Traditionally, the gap or ``green valley" between these two populations has been interpreted as evidence that galaxies undergo a rapid transition from star-forming to completely quiescent, forming a sparsely populated region in this parameter space \citep{bgb2004, bwm2004, fww2007, thb2015}. However, more recent studies have proposed various quenching timescales \citep{mws2007, sus2014, wdf2015, bfk2015}, where galaxies that quench slowly may account for a large fraction of the green valley population. As such, understanding the physical mechanism(s) behind how individual galaxies transform and traverse through the ``green valley" in order to produce the growth of the quiescent population has been a major topic of research in extragalactic astrophysics.

Star formation requires gas cooling down to a cold molecular form before clumps and cores begin to form systems of stellar nurseries. Therefore, the physical mechanism producing quiescence must somehow either stop gas from cooling to this form or eject the gas completely for extended periods of time. In this paper, we focus on the physics behind quiescence in central galaxies, or those at the centers of their dark matter halos, since satellite galaxies undergo unique processes that are specific to systems located well within the hot, gaseous atmospheres of other galaxies. 

A multitude of possible mechanisms affecting central galaxies have been proposed: stellar and supernovae Ia feedback \citep{wr1978, ds1986, wf1991, hqm2012}, halo mass quenching \citep{db2006, cdd2006, bdn2007, dbe2009, gd2015}, morphological quenching \citep{kk2004, mbt2009, cjb2011}, stellar mass quenching \citep{plk2010}, gravitational heating \citep{jno2009}, and varying forms of black hole feedback from active galactic nuclei \citep[AGN,][]{kh2000,dsh2005, csw2006, cfb2009, f2012, cms2014}.


\begin{figure*}
\centering
\epsfxsize=18cm
\epsfbox{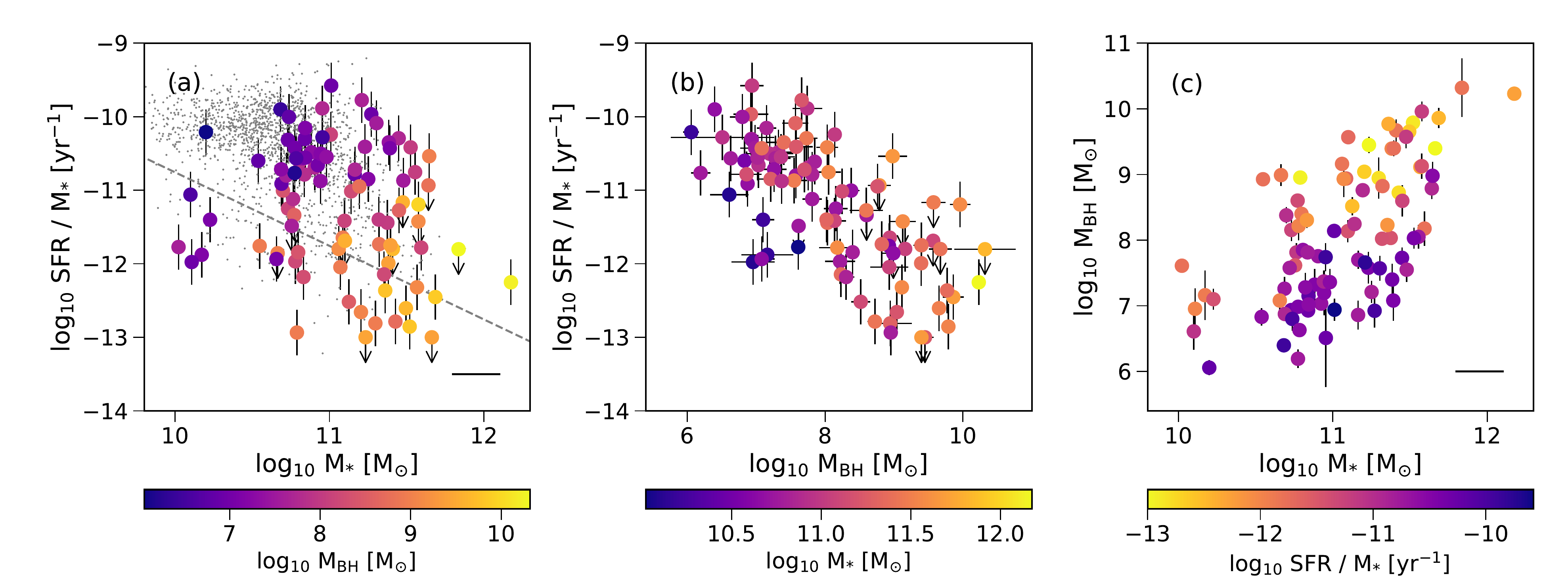}
\caption{Projections of the 3-dimensional sSFR--$M_{\rm{BH}}$--$M_{*}$ data cube: (a) sSFR as a function of $M_{*}$. The gray data points indicate a sample of local galaxies to show the star forming main sequence. The dashed line indicates the boundary below which the sample is no longer complete; (b) sSFR as a function of $M_{\rm{BH}}$; (c) $M_{\rm{BH}}$ as a function of $M_{*}$. Color gradients indicate the values for the axis not shown. The lines at the bottom right of (a) and (c) indicate the errors on $M_{*}$.}
\label{fig:3Dplots}
\end{figure*}

Observationally, quiescent galaxies are more common with increasing stellar mass, and tend to host massive bulges, concentrated central stellar surface densities, concentrated light profiles, higher central velocity dispersions, and more massive dark matter halos \citep{khw2003, fvf2008, bvp2012, lws2014, bme2014, wdf2015, mwz2016}. Many of these properties are expected to correlate closely with the central supermassive black hole mass \citep{kh2013}, lending support to the idea that black hole-driven feedback is important for producing quiescence in central galaxies.

Recently, a myriad of studies have compiled an ever-growing list of dynamically-measured black hole masses \citep[e.g.,][]{kh2013, soe2016, v2016}, allowing a more direct and statistical study of how black holes and galactic properties correlate with one another. \citet{tbh2016b} used a combination of these compilations in order to show that quiescent galaxies have more massive black holes than star-forming galaxies at a given stellar mass. They also show that this behavior is naturally produced in models where star formation is regulated by long-lived radio-mode AGN feedback. 

This paper aims at expanding their study by exploring exactly how the star formation rate of a galaxy correlates with its black hole mass and stellar mass, thereby probing the way in which the black hole responsible for AGN feedback affects the amount of star formation occurring in the galaxy.

We begin by presenting the galaxy data we use in our analysis ($\S$2) and go on to describe the resulting trends and correlations produced by the data ($\S$3). We then discuss the physical framework we propose in order to interpret our results in the context of AGN feedback ($\S$4.1). This motivates a discussion on whether galaxies which host intermediate amounts of star formation, or what we call `partially quiescent' galaxies, are truly transitioning ($\S$4.2). Model results are then shown in order to compare our physical interpretation with the results from detailed simulations of galaxy formation ($\S$4.3). Finally, we end with concluding remarks ($\S$5).



\begin{figure*}
\centering
\epsfxsize=18cm
\epsfbox{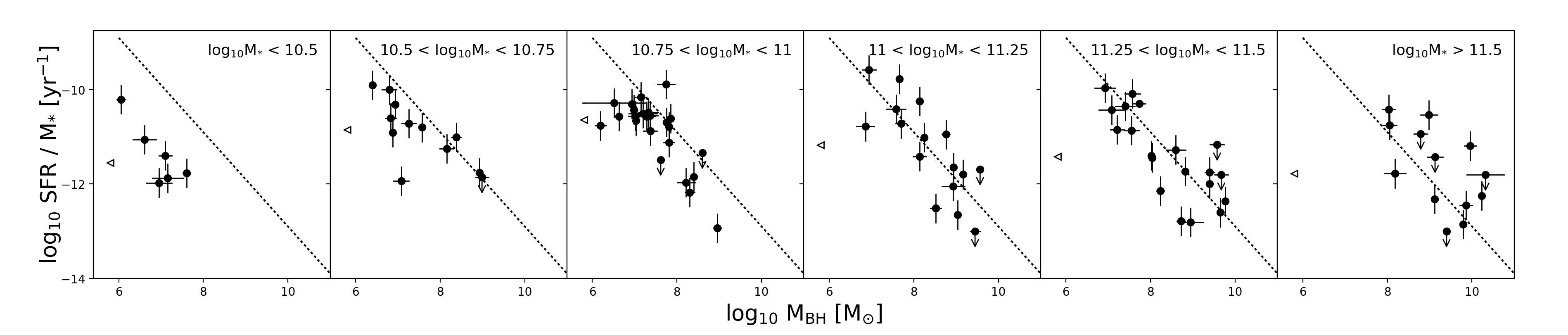}
\caption{sSFR as a function of $M_{\rm{BH}}$ for different bins of $M_{*}$. The dotted lines are the same in each panel in order to compare the relations at high and low $M_{*}$ bins. Open, left-facing triangles indicate the median sSFR at each $M_{*}$ bin.}
\label{fig:sSFR_Mbh}
\end{figure*}

\section{Data}
\label{sec:obsdata}

We adopt the sample of nearby (z $\lesssim$ 0.034 or d $\lesssim$ 150 Mpc) galaxies with dynamical estimates of black hole masses ($M_{\rm{BH}}$) from \citet{tbh2016b}, where the base sample comes from \citet{soe2016} and is supplemented by \citet[and references therein]{v2016}. Our sample selects only central galaxies, identified as the brightest or only members in their association within a $\sim$1 Mpc radius in order to focus on the physics of quiescence for galaxies at the centers of their potential wells.

We use extinction-corrected 2MASS `total' $K_{\rm{s}}$ apparent magnitudes \citep{hmm2012} to infer galaxy stellar masses ($M_{*}$), adopting a single K-band stellar $M_{*}/L_{\rm{K}}$ ratio of 0.75 and assuming an uncertainty of 0.15 dex. In order to calculate star formation rates (SFRs), we use far-infrared (FIR) fluxes obtained by IRAS \citep[][see also corrections to \citealp{kgk1989} in NED by Knapp 1994]{rls1988, m1990, ssm2004, sh2009} in conjunction with the methods to derive SFR described in \citet{ke2012}. Galaxies with no infrared detections or detections that result in SFR/$M_{*}$ $<$ 10$^{-13}$ yr$^{-1}$ are shown as upper limits. We adopt a 0.3 dex uncertainty for our SFR values \citep{b2003}. Refer to \citet{tbh2016b} for more information on the methods for calculating galaxy properties.

Table~\ref{table1} shows the data we used in our study and Table~\ref{table2} shows the data we used to infer galaxy properties from the literature.

\section{Results}
\label{sec:results}

Figure~\ref{fig:3Dplots} shows the sSFR--$M_{*}$, sSFR--$M_{\rm{BH}}$, and $M_{\rm{BH}}$--$M_{*}$ parameter space for our sample of 91 central galaxies. These plots show three projections of a 3-dimensional data cube where the color gradient in each panel represents the values of the axis not shown. We can see a clear correlation between sSFR, $M_{\rm{BH}}$, and $M_{*}$ - namely, for a given $M_{*}$, quiescent galaxies have more massive $M_{\rm{BH}}$ than star-forming galaxies, as is shown and discussed in \citet{tbh2016b}. However, galaxies at a given $M_{*}$ can have diverse sSFRs which generally decrease with increasing $M_{\rm{BH}}$ as can be seen in the color gradient in the rightmost panel of Figure~\ref{fig:3Dplots}. This trend appears to be continuous in our data, motivating us to avoid classifying galaxies into two broad categories of `star-forming' and `quiescent.' Thus, we choose to explicitly explore whether the sSFR distribution produces a dichotomy or instead varies more continuously as a function of other galaxy parameters.


In order to investigate this, we focus on the central panel of Figure~\ref{fig:3Dplots} and show sSFR as a function of $M_{\rm{BH}}$ separated into bins of $M_{*}$ in Figure~\ref{fig:sSFR_Mbh}. We find that the sSFR is a smoothly declining function of $M_{\rm{BH}}$ in each $M_{*}$ bin. The dotted black line is the same in all panels in order to show the similar slope of the relation at all $M_{*}$ bins. There is also an offset in the relation between different $M_{*}$ bins where less massive galaxies tend to have lower sSFRs at a given $M_{\rm{BH}}$ than more massive galaxies. We also note that while there is a wide range of sSFRs at a given $M_{*}$, the median sSFR at each $M_{*}$ bin (open, left-facing triangles) gradually decreases as $M_{*}$ increases at log$_{10} M_{*} > 10.75$, in accordance with the observation that more massive galaxies tend to be more quiescent. The galaxies detected at the two lowest $M_{*}$ bins show lower median sSFRs. This is likely due to the fact that these galaxies are not representative of the general galaxy population at these $M_{*}$ bins, since most low mass galaxies probably have central black holes with masses too low to be detected \citep{rgg2013}. Finally, we note that more massive galaxies tend to have more massive black holes although the scatter is substantial as is evident in the right panel of Figure~\ref{fig:3Dplots}.

The presence of a vertical offset for different $M_{*}$ bins for the relations shown in Figure~\ref{fig:sSFR_Mbh} hints at the fact that galaxies form a manifold in this three dimensional space. We choose to fit the simplest three-dimensional manifold -- a plane -- to the sSFR--$M_{\rm{BH}}$--$M_{*}$ distribution for our sample of galaxies using a linear ordinary least squares analysis, excluding galaxies with sSFR upper limits from our fit. The result is described by the equation: 

\begin{equation}
\begin{aligned}
\rm{log}_{10} sSFR = (0.80 \pm 0.18) \ log_{10} \frac{\textit{M}_{*}}{\textit{M}_{*,\rm{avg}}} \\ - \ (0.82 \pm 0.08) \ \rm{log}_{10} \frac{\textit{M}_{\rm{BH}}}{\textit{M}_{\rm{BH},\rm{avg}}} \\ - \ (11.84 \pm 0.10),
\end{aligned}
\end{equation}
where we adopt bootstrap errors. We normalize $M_{*}$ and $M_{\rm{BH}}$ by their average values for our sample, where $M_{*,\rm{avg}}$ = 1.62$\times$10$^{11}$ M$_{\odot}$ and $M_{\rm{BH},\rm{avg}}$ = 8.71$\times$10$^{8}$ M$_{\odot}$.

\begin{figure}
\epsfxsize=8.5cm
\epsfbox{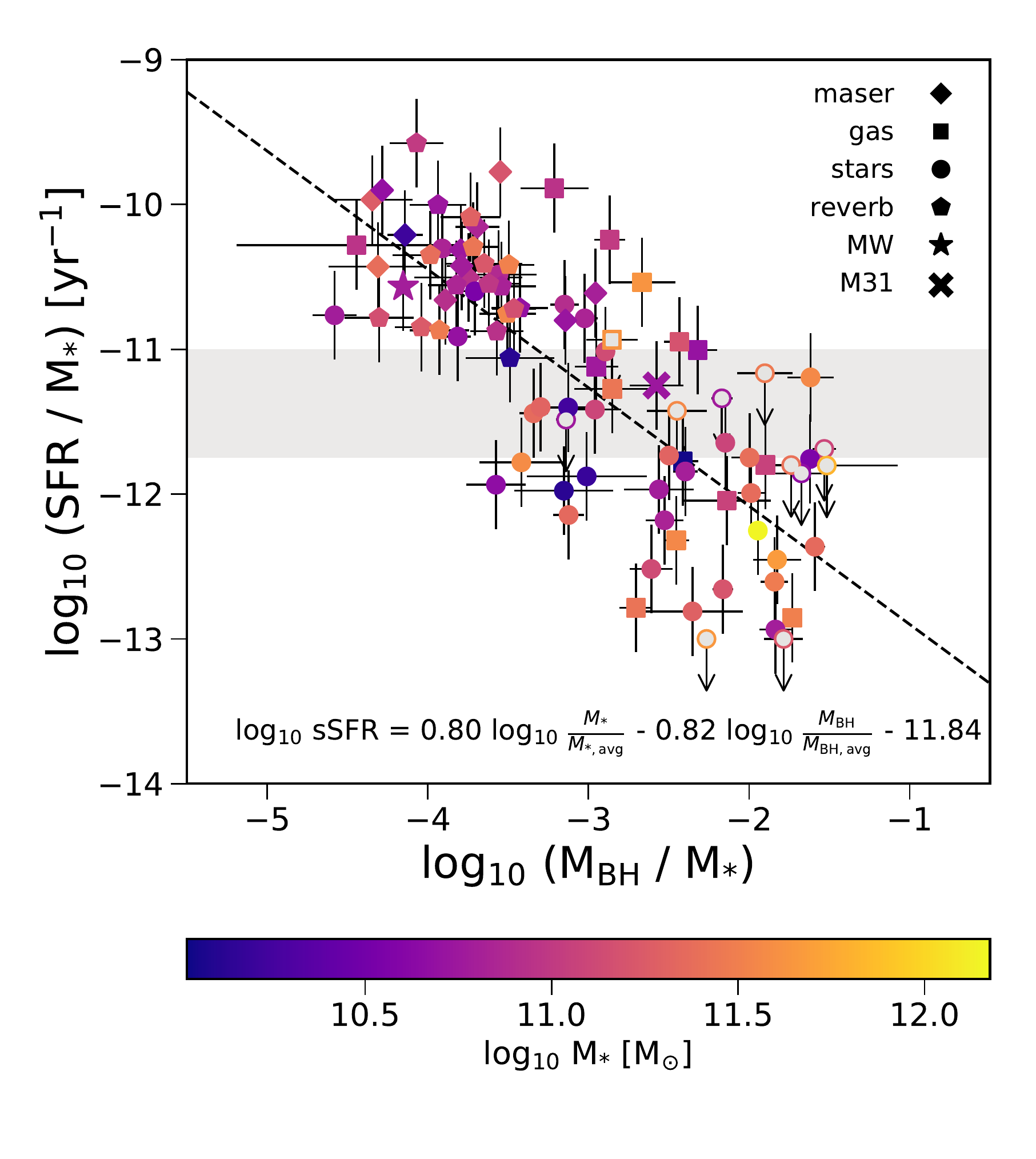}
\caption{sSFR as a function of $M_{\rm{BH}}$/$M_{*}$ or $sM_{\rm{BH}}$ for our sample. The dashed line indicates the best fit plane described by the equation shown at the bottom of the panel. Upper limits are not included in the fit and are indicated by open, unfilled data points. The light gray shaded region highlights galaxies which are partially quiescent (See Section~\ref{sec:partialq}). The color gradient indicates $M_{*}$.}
\label{fig:partialq}
\end{figure}

We note that the powers for $M_{*}$ and $M_{\rm{BH}}$ are about equal. For this reason, Figure~\ref{fig:partialq} shows a projection of this plane by plotting log$_{10}$sSFR against the logarithm of the ratio between $M_{\rm{BH}}$ and $M_{*}$, or what we will call the galaxy's specific black hole mass ($sM_{\rm{BH}}$). Dividing by $M_{*}$ effectively reduces the $M_{*}$ dependence the sSFR has on the $M_{\rm{BH}}$. We find that the sSFR is a smoothly decreasing function of the $sM_{\rm{BH}}$ for the overall population where the scatter is $\sim$0.55 dex. Our result applies to a diversity of galaxy types, ranging from disky to spheroidal structures and spans a range of four orders of magnitude in sSFR, two orders of magnitude in stellar mass, and five orders of magnitude in black hole mass.

We color the data points by $M_{*}$ to show two important features. First, we note that $M_{*}$ and sSFR correlate poorly with one another compared to the correlation between $sM_{\rm{BH}}$ and sSFR. In other words, galaxies with similar $M_{*}$ can be found anywhere along the relation, with any sSFR value, since they can have a wide variety of $sM_{\rm{BH}}$ values. Second, while this first point is true, more massive galaxies tend to preferentially have larger $sM_{\rm{BH}}$ and lower sSFRs while the opposite is true for less massive galaxies. This reflects the general trend that more massive galaxies tend to host less star formation while potentially hinting at the source of scatter in sSFR at a given $M_{*}$.

We note that a similar negative correlation is found in the central panel of Figure~\ref{fig:3Dplots} where there is no dependence on $M_{*}$. We find that the scatter in sSFR at a given $M_{\rm{BH}}$ is 0.61 dex. Allowing the sSFR to be a function of both $M_{\rm{BH}}$ and $M_{*}$ provides a better fit with 0.55 dex scatter -- corresponding to a reduction by a fifth of the total variance in the central panel of Figure~\ref{fig:3Dplots} -- and is preferred at $>$99.99 percent level of all fits of bootstrapped samples having no $M_{*}$ dependence. Even so, we stress that the exact powers of $M_{\rm{BH}}$ and $M_{*}$ need to be confirmed with a larger and more complete sample than what current black hole data sets offer. We note that alternative versions of this fit, using different prescriptions for estimating SFR (e.g., including UV detections) and different selections for the central galaxy sample give similar results, with the fit parameters varying within their quoted errors.

\section{Discussion}

\subsection{Physical Framework and Interpretation}
\label{sec:physics}

Our main result is that the sSFR of a galaxy correlates smoothly with $sM_{\rm{BH}}$, suggesting that the star-forming properties of a galaxy are somehow aware of the properties of the central black hole. While the amount of scatter is significant at 0.55 dex, the negative correlation in our data is clearly present. In order to make physical sense of this, we argue that the sSFR can only know about the $M_{\rm{BH}}$ and $M_{*}$ if one of two scenarios are occurring: (1) black hole feedback, assuming it is measurable via $M_{\rm{BH}}$, is regulating the amount of star formation in the galaxy to some degree, or (2) the increase in $M_{\rm{BH}}$ and decrease in sSFR are due to a strongly correlated but separate process where there is no direct causal connection between the two.

Recent galaxy formation models have relied on black hole-driven AGN feedback as the primary cause of quiescence \citep{csw2006, ssd2007, gwl2010, vgs2014, psp2014, hwt2015, scb2015} since no other mechanism can produce a strong enough suppression of stellar mass build up in high mass galaxies \citep{bbm2006}. \citet{tbh2016b} shows that out of the four models analyzed, only those models that use radio-mode AGN feedback to provide a continuous source of heat reproduce the observational result that quiescent and star-forming galaxies lie on distinct regions on the $M_{\rm{BH}}$-$M_{*}$ plane where quiescent galaxies have more massive black holes than star-forming galaxies; this is unlike those models that use halo mass quenching or quasar-mode AGN feedback as the primary source of quiescence. As a result, we will set up a physical framework where we focus on scenarios which causally link the $M_{\rm{BH}}$ and the sSFR of a galaxy.


Successful simulations have modeled AGN feedback in the radio mode as bubbles expanding into the circumgalactic medium around a galaxy in order to heat the surrounding gas, thereby cutting off the fuel needed for star formation \citep[e.g.,][]{csw2006,ssd2007}. While black hole feeding likely happens at irregular intervals depending on gas availability, bubbles formed by radio-mode AGN feedback are expanding into the medium long after accretion stops. In this physical scenario, the heating from the expansion of these bubbles is likely to be more or less continuous even though black hole feeding is not. Observationally, this idea is supported by the presence of long-lived X-ray cavities and `ghost' cavities from past accretion events in the intracluster medium around cluster, group, and isolated galaxies \citep{brm2004, djf2009, gog2010, swm2016}.

Models also show that the $M_{\rm{BH}}$ correlates with the amount of heating from AGN feedback \citep{svg2015,tbh2016b}. In accordance with this, our observational results in Section~\ref{sec:results} show that larger values of $sM_{\rm{BH}}$ result in correspondingly lower values of sSFR to produce a negative correlation. A possible interpretation is that the sSFR adjusts to the $sM_{\rm{BH}}$ at least at $z = 0$ to produce a smoothly declining relation between these two parameters. This adjustment must happen on short enough time scales to allow such a relation between sSFR and $sM_{\rm{BH}}$. In other words, if either of these quantities could drastically change without allowing the other quantity to adjust, then a relation between these two parameters would not appear as clearly as it does in Figure~\ref{fig:partialq}.



In addition, we note that the shape of the relation is important: a smoothly declining relation may hint at the physics behind heating and cooling of gas around the galaxy. More specifically, this may mean that an intermediate $sM_{\rm{BH}}$ results in an intermediate amount of gas heating which decreases, but does not completely halt, the amount of gas cooling onto the disk to fuel star formation in the galaxy. We expand on this issue in Section~\ref{sec:partialq} where we discuss the phenomenon of partial quiescence.

The vertical offset in the relations between sSFR and $M_{\rm{BH}}$ from low to high $M_{*}$ bins in Figure~\ref{fig:sSFR_Mbh} can be interpreted to mean that more massive galaxies need a more massive black hole to maintain the same degree of quiescence as less massive galaxies, since more massive galaxies have a deeper potential well and, in the absence of heating, would be forming more stars as a result of cooling and gravity. However, more massive galaxies also tend to have lower sSFRs than the less massive galaxies, in general agreement with other studies \citep{bcw2004, src2007}. This implies that the $M_{\rm{BH}}$ of massive galaxies are significantly larger than those of less massive galaxies, resulting in the vast majority of the high $M_{*}$ galaxy population to be predominantly quiescent. In addition, the fact that we normalize both the SFR and $M_{\rm{BH}}$ by $M_{*}$ tells us that reducing the dependence on the depth of the potential well -- represented by $M_{*}$ in this paper -- gives us a similar relation across a diverse group of central galaxies with $M_{*} > 10^{10}$ M$_{\odot}$.

We also note that the scatter between sSFR and $M_{\rm{BH}}$ increases in the highest $M_{*}$ bin in Figure~\ref{fig:sSFR_Mbh}. This could be due to multiple factors. For one, $M_{*}$ is likely an increasingly poor tracer of a galaxy's potential well at high $M_{*}$ since the $M_{*}$--$M_{\rm{h}}$ relation becomes substantially flatter at these mass regimes \citep{msm2010, bcw2010, bwc2013}. Instead, obtaining a halo mass may be more effective, albeit more difficult, and may eliminate the increased scatter. In addition, low sSFR values are increasingly difficult to measure and may have less meaning with regards to the actual amount of star formation in the galaxy. Another explanation may be that more massive galaxies are probing clusters rather than groups and isolated galaxies. Black hole feedback in these systems may differ in terms of how gas heating and cooling operates \citep{gbd2011}.

An important assumption we have made is that $M_{\rm{BH}}$ is measuring the amount of heating energy being injected into the gas around the galaxy while the sSFR is measuring the amount of gas cooling onto the galaxy. In the real Universe, these parameters may vary widely on a galaxy-to-galaxy basis based on the state of the gas, the star formation efficiency of the galaxy, the duty cycle and jet power of the black hole feedback and how that correlates with $M_{\rm{BH}}$, and potentially many other factors.


\subsection{Partial Quiescence}
\label{sec:partialq}

Galaxies which have low yet significant amounts of star formation in our sample are shown in the light gray band in Figure~\ref{fig:partialq}. Previous studies have often referred to these galaxies as transitioning or ``green valley" galaxies \citep{bwm2004, bbn2004, mws2007, bwv2009, mcl2011, wtc2012, gmm2012, khm2013, pll2014}, assuming they are on their way to becoming completely quiescent. Given the dearth of galaxies in this region on a color-magnitude diagram, the traditional view is that galaxies quickly move from the blue cloud to the red sequence or, as has been interpreted, from star-forming to quiescent \citep[e.g.,][]{bgb2004}.




A smoothly decreasing correlation between the sSFR and $sM_{\rm{BH}}$ shown in Figure~\ref{fig:partialq} and described in Section~\ref{sec:results} is perhaps unexpected given the commonly held belief that galaxies exist only briefly in this transition state. If the $M_{\rm{BH}}$ grows significantly, then in order to land on the relation in Figure~\ref{fig:partialq} and agree with our observational result, the galaxy must also decrease its sSFR accordingly. Therefore, a star-forming galaxy that grows its black hole to an intermediate $sM_{\rm{BH}}$ must also decrease its sSFR to an intermediate value on timescales short enough to produce a relation between the two quantities. 

We note that this framework \textit{does not require central galaxies with intermediate sSFRs to be transitioning at all}. A central galaxy can stay in the ``green valley" as long as it no longer grows its $sM_{\rm{BH}}$. In this scenario, the relation between sSFR and $sM_{\rm{BH}}$ represents the amount of star formation that results from the balance between heating and cooling represented by the ratio between a galaxy's $M_{\rm{BH}}$ and $M_{*}$. If this is the case, then this framework implies that all central, massive galaxies tend towards an equilibrium position defined by this relation which determines their sSFR from their $sM_{\rm{BH}}$, and that much of the scatter likely comes from the time it takes for the sSFR to adjust to the $sM_{\rm{BH}}$.


As a result, the fact that the sSFR, $M_{\rm{BH}}$, and $M_{*}$ are smooth but scattered functions of each other leads us to argue that many of the partially quiescent galaxies in our sample may not be transitioning -- instead they may maintain a quasi-stable state of quiescence that correlates with their $M_{\rm{BH}}$ and $M_{*}$.


One possible example of this in our sample is M31, labeled in Figure~\ref{fig:partialq}. M31 is not undergoing any drastic event that suggests it is quenching and heading towards a completely quiescent state. Yet many studies have shown that M31 has a lower sSFR \citep{kbr2009, fgs2013, ldd2015} than expected based on where a galaxy with its stellar mass would be if it were on the star forming main sequence. In our framework, this simply comes from the fact that M31 has an over-massive black hole for its stellar mass (i.e. a higher $sM_{\rm{BH}}$) and as a result gives us a lower sSFR. Similarly, M81 also lands within the partially quiescent sample and does not exhibit any morphological signs of transitioning.


If there are a significant number of stable galaxies with intermediate sSFRs, then this implies a more populated ``green valley" than previously observed. In support of this implication, \citet{oag2016} argue that the ``green valley" is more populated than is otherwise believed due to the selection effects, systematic errors, and bias they find in one of the more popular collections of SFRs from SDSS \citep{bcw2004}. They present a representative sample of local galaxies with updated and reliably-measured SFRs from ultraviolet and mid-IR fluxes and find a significantly larger, distinct population of galaxies with intermediate sSFRs. In addition, \citet{edw2017} argue that the galaxy population exhibits a gradual difference in properties between star-forming and quiescent galaxies, a behavior that is erased in color-space due to colors varying minimally below a threshold value of sSFR. This would challenge the widely accepted view that galaxies live in two distinct populations, and instead argue for a more unitary approach. Finally, many studies have also argued for the existence of varying degrees of quiescence that could hint at a class of galaxies that spend an extended amount of time in the ``green valley" \citep[e.g.,][]{lyz2016, pbs2016}. Even so, our proposed framework of sSFR regulation by the black hole does not necessarily require a continuous distribution of galaxies along this relation since this distribution depends strongly on the details of black hole growth.

We note that our sample selection is biased and heterogeneous due to our requirement of a dynamical black hole mass measurement using a variety of detection methods. This impacts our analysis in two ways. First, while we do not detect a significantly underpopulated ``green valley" for the data in our sample, the current black hole data available are insufficient to probe the prominence of the ``green valley" in the general central galaxy population since the sample is not representative. Future work will be important for determining the prominence of the ``green valley" and the strength of bimodality in sSFR parameter space for central galaxies at this mass regime using reliable SFR indicators (See \citealt{oag2016} and \citealt{edw2017} for important progress). Second, while it is clear that sSFR is a smoothly decreasing function of $sM_{\rm{BH}}$ for the current black hole data for central galaxies, we caution that the exact form of the relation may be impacted by selection. For example, studies using the central stellar mass density within 1 kiloparsec ($\Sigma_{1 \rm{kpc}}$) as an indirect proxy for $M_{\rm{BH}}$ see a relatively sudden drop in sSFR as a function of $\Sigma_{1 \rm{kpc}}$ \citep{ffk2013, wdf2015}. We do not see evidence of this sudden drop in sSFR with our dataset, perhaps due to the inadequacy of such proxies for $M_{\rm{BH}}$ or due to our sample's size and inhomogeneities. It would be important to quantify the degree to which the dependence of sSFR is gradual and continuous with a larger and more complete sample, and to remain open to any higher order structure in the sSFR--$M_{\rm{BH}}$--$M_{*}$ parameter space.


While we are proposing the possibility that much of the ``green valley" population is in a quasi-stable state of partial quiescence, we also recognize that there likely exist various pathways a galaxy could take as it grows its black hole and stellar mass and varies its star formation rate. For example, rapid and more violent processes perhaps more common for giant ellipticals may skew the observed relationship between the sSFR, $M_{\rm{BH}}$, and $M_{*}$ as may be shown in the increased scatter in the relation between $M_{\rm{BH}}$ and sSFR at high stellar masses. The existence of more than one quenching mode and speed has been discussed in other works \citep{mws2007, bfp2013, sus2014, wdf2015, lyz2016}. Whereas a quasi-stable state of partial quiescence would be consistent with slow quenching since the relevant timescales are comparable to or longer than a Hubble time, those experiencing faster quenching would likely account for some of the scatter between sSFR and $sM_{\rm{BH}}$. In addition, other processes that affect a galaxy's sSFR such as morphological quenching, stellar and supernovae Ia feedback, merging, or gravitational heating may affect a galaxy's position on the sSFR--$sM_{\rm{BH}}$ plane. Even so, the clear correlation between these three parameters shows that, if our physical framework is at least generally valid, black hole feedback is the most important physical mechanism in determining a galaxy's star formation properties and that a large part of the galaxy sample can be characterized as being in a quasi-steady state or approaching this state as the sSFR responds to the change in $sM_{\rm{BH}}$ that the galaxy has undergone.



\begin{figure*}
\epsfxsize=12cm
\centering
\epsfbox{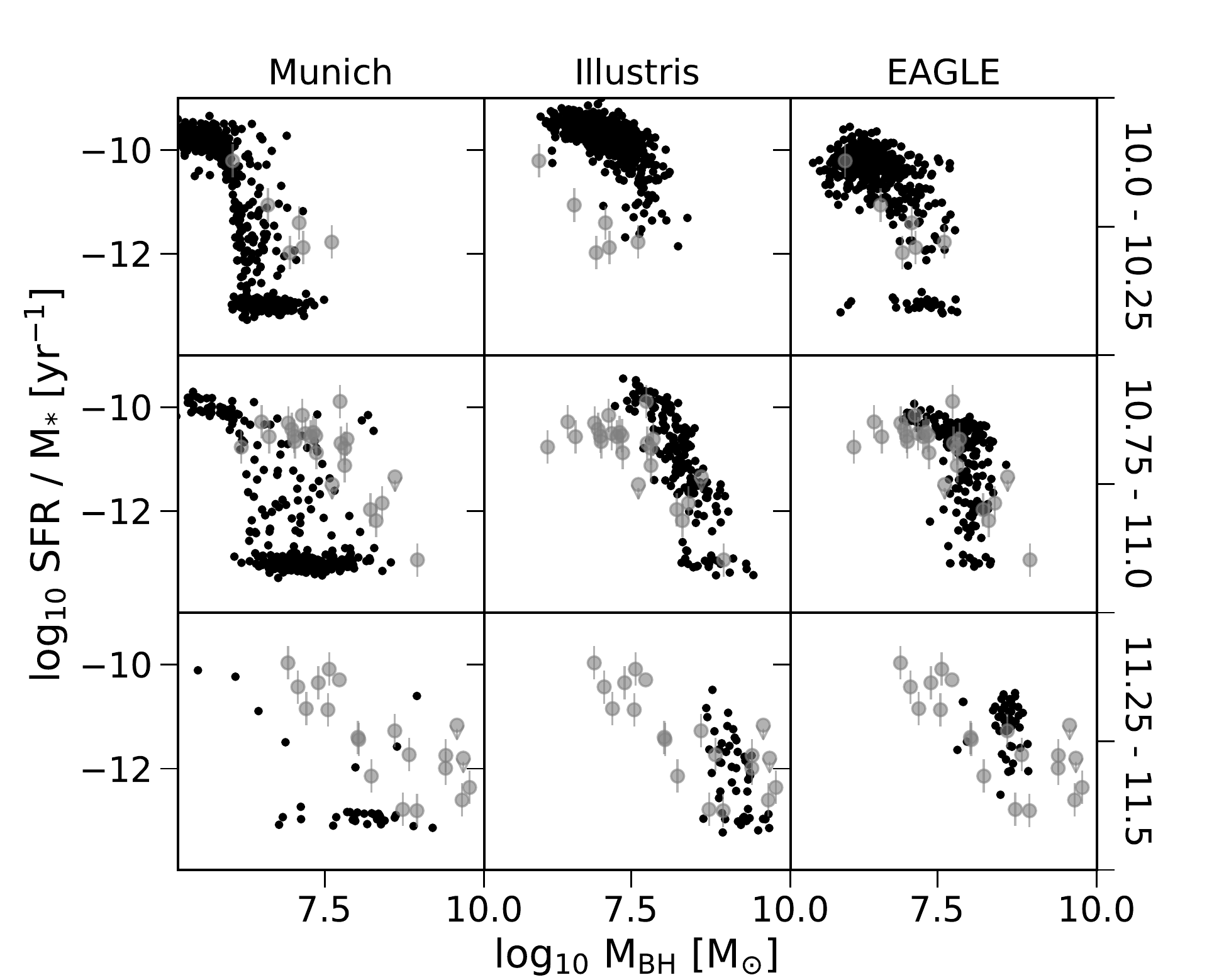}
\caption{sSFR as a function of $M_{\rm{BH}}$ in three bins of $M_{*}$ for the Munich, Illustris, and EAGLE galaxy formation models (black points). Overlaid is the observational data from Figure~\ref{fig:sSFR_Mbh} in the corresponding $M_{*}$ bins (gray translucent points). Each column indicates different models and each row indicates the $M_{*}$ bin: 10.25 $<$ log$_{10}$ $M_{*}$ $<$ 10.5 (top); 10.75 $<$ log$_{10}$ $M_{*}$ $<$ 11.0 (middle); and 11.25 $<$ log$_{10}$ $M_{*}$ $<$ 11.5 (bottom).}
\label{fig:models}
\end{figure*}

\subsection{Model Comparison}
\label{sec:models}

\citet{tbh2016b} shows a strong correlation between $M_{\rm{BH}}$ and quiescence at a given stellar mass. When comparing these results to state-of-the-art models, they found that the latest Munich semi-analytic model \citep{hwt2015} and the Illustris hydrodynamic simulation \citep{vgs2014} showed the best agreement with observations unlike the EAGLE hydrodynamic simulation \citep{scb2015} and GalICS semi-analytic model \citep{cdd2006}. Here we focus on the Munich, Illustris, and EAGLE simulations since these explicitly use AGN feedback as the primary mechanism behind quiescence.

The Munich model includes a continuous heating rate affecting the temperature of the circumgalactic medium that depends on the hot halo gas mass and $M_{\rm{BH}}$. Illustris introduces buoyant bubbles which expand into the atmosphere every time the black hole is fed cold gas. While the creation of these bubbles is stochastic, the effect they have on the temperature of the circumgalactic medium is gradual as the bubble slowly expands into the gas. Hence, both of these models use either continuous or quasi-continuous heating from radio-mode AGN feedback in order to shut off star formation in galaxies at the high end of the stellar mass function. In contrast, EAGLE uses quasar-mode feedback to intermittently inject energy into the interstellar medium only when there is gas available to the black hole.

We find that the quantitative relationship between sSFR, $M_{\rm{BH}}$, and $M_{*}$ vary from model to model. For example, at the stellar mass regimes of interest, the Munich model's determination of a central galaxy's sSFR has little to no dependence on the stellar mass of the central galaxy and instead depends strongly on a $M_{\rm{BH}}$ threshold, see \citealp{tbh2016a}. Conversely, Illustris's sSFRs depend more strongly on $M_{*}$ such that more massive galaxies need more massive black holes in order to have the same sSFR as a lower mass galaxy, see \citealp{tbh2016b}. In Section~\ref{sec:results} we fit a plane to our observational data which demonstrated that \textit{in the real Universe} sSFR is a smoothly decreasing function of the ratio between $M_{\rm{BH}}$ and $M_{*}$ (what we are calling the $sM_{\rm{BH}}$). However, in the Munich model, for example, a ratio of $M_{\rm{BH}}$ and $M_{*}^{0.1}$ would better reveal the physics behind quiescence since the Munich model's sSFRs barely depend on $M_{*}$ and therefore requires $M_{*}$ to have a smaller power. As a result, the $sM_{\rm{BH}}$ will not be useful for understanding the suppression of sSFR in the models since they do not agree with observations in this respect.

Rather than introducing different powers of $M_{*}$, we choose to compare the models to our observational results by presenting the sSFR--$M_{\rm{BH}}$ plane at different $M_{*}$ bins for each of these models. This effectively focuses on the dependence between the sSFR and $M_{\rm{BH}}$ rather than the differences between $M_{*}$ dependencies in the models. We show the distributions of these galaxies in Figure~\ref{fig:models} where for each model we select only central galaxies within a 100 Mpc$^{3}$ volume at $z=0$. The $M_{*}$ bins are directly comparable to those in the first, third, and fifth panels of Figure~\ref{fig:sSFR_Mbh} whose data points are overplotted in gray. Any galaxies with sSFR $< 10^{-13}$ yr$^{-1}$ are assigned an arbitrarily low sSFR value defined by a normal distribution around this limit.

We find that the Munich model (left panels) exhibits a steep drop off in sSFR at a given $M_{\rm{BH}}$ for most central galaxies. A clear bimodality exists where galaxies either have high or low sSFRs with a few galaxies in between. In the highest $M_{*}$ bin there are very few galaxies with most of them having a massive black hole and therefore having very low sSFR values. \citet{tbh2016a} show that most galaxies in this model are immediately quenched as soon as they reach a redshift-dependent critical $M_{\rm{BH}}$ (See their Section 3 for more details).

The Illustris simulation (center panels) shows a considerably different distribution on this plane. Rather than showing a steep drop in specific star formation rate as a function of $M_{\rm{BH}}$ at a given $M_{*}$, it shows a smoothly declining function for galaxies with massive enough black holes to begin suppressing star formation, much like what is seen in our observational results in Figure~\ref{fig:sSFR_Mbh}. It is clear, however, that black hole mass correlates more tightly with stellar mass in Illustris than in our observational sample since there there is a larger variety of black hole masses at each stellar mass bin in Figure~\ref{fig:sSFR_Mbh} than for Illustris. In addition, galaxies with $M_{\rm{BH}} \lesssim 10^{7}$ M$_{\odot}$ do not exhibit much dependence on $sM_{\rm{BH}}$ since these galaxies' sSFRs are likely not regulated by AGN feedback in this model.

The EAGLE simulation (right panels) exhibits an L-shaped distribution where galaxies are mostly star-forming until they reach a certain $M_{\rm{BH}}$ value depending on their $M_{*}$, where many but not all galaxies begin to have lower sSFRs. As discussed in \citet{tbh2016b}, this behavior is not reflected in the observation that star-forming and quiescent galaxies have distinct black hole mass distributions at a given $M_{*}$. The overlap in these distributions at high $M_{\rm{BH}}$ in this model is likely due to the fact that galaxies in EAGLE undergo intermittent heating episodes rather than a continuous injection of energy. In this model, galaxies can continue forming stars once again in between feedback events even with a massive central black hole, producing an L-shaped distribution that is not reflected in our observational results. In other words, in EAGLE, a massive black hole is a necessary but not sufficient condition for quiescence in central galaxies since star-forming galaxies can also host massive black holes.

Each of these models have been quite successful in reproducing many of the observational trends out to $z = 2$, particularly the Munich model. Even so, it is clear that different physical implementations of quiescence can drastically affect the distributions in the sSFR--$M_{\rm{BH}}$--$M_{*}$ parameter space, even when they use similar physical frameworks for AGN feedback. As such, none of the models match our observational results perfectly as is clear from the overlaid observational data (gray translucent points) in each panel. The models differ from each other and from the observations in this parameter space with respect to their variety of dependences on $M_{*}$, black hole mass distributions, stellar mass distributions, and strength of bimodality. This manifests itself as differences in the shape of the distributions, the slopes of the decline in sSFR, the normalization of the distributions in each $M_{*}$ bin, and the scatter of sSFR as a function of $M_{\rm{BH}}$.

However, qualitatively, we find that the results from Illustris better resemble our observational results. We note that the smoothly-declining yet scattered relation between sSFR and $M_{\rm{BH}}$ in Illustris shows that even in an idealized simulation, an appreciable amount of scatter, such as what is seen in our observations, is expected within a framework where AGN feedback determines a galaxy's star formation properties.

Even so, it is well-established that the AGN feedback in Illustris is too violent and ejects too much gas out of its hot halo \citep{vgs2014}. Further implementations of Illustris must be tested in order to understand whether this behavior persists with a less violent AGN feedback model. By extension, future models will need to consider how their prescriptions for AGN feedback correspond to the largely unexplored idea of a smoothly decreasing correlation between sSFR and $sM_{\rm{BH}}$ along with the idea of partial quiescence.


\section{Conclusions}
\label{sec:conclusions}

In order to more directly and statistically study AGN feedback in the context of galaxy relations, we choose to study the correlation between a galaxy's sSFR, $M_{\rm{BH}}$, and $M_{*}$. We have shown that for our diverse sample of 91 central galaxies with dynamically detected $M_{\rm{BH}}$, sSFR is a smoothly decreasing function of $M_{\rm{BH}}$/$M_{*}$, or what we call the specific black hole mass, $sM_{\rm{BH}}$. In an attempt to interpret this correlation, we propose a physical framework where the amount of gas heating from radio-mode AGN feedback is reflected by $M_{\rm{BH}}$ and combats the supply of fuel for star formation within the galaxy. In this framework, a galaxy with a more massive $sM_{\rm{BH}}$ would have a correspondingly lower sSFR, in accordance with our observational result. 

This framework provides an alternative to the idea that all ``green valley" galaxies are transitioning from star-forming to quiescent phases. Instead, it predicts that these galaxies with intermediate values of $sM_{\rm{BH}}$ live in a stable state of partial quiescence between star-forming and quiescent galaxies. 

No current models achieve the distribution of galaxies that we see in this three-dimensional parameter space, although Illustris comes close. Future work will need to take these observational constraints into account when implementing AGN feedback models in order to shut off star formation in central galaxies.

\acknowledgements

B.A.T. is supported by the National Science Foundation Graduate Research Fellowship under Grant No. DGE1256260. We acknowledge helpful discussions with S. White, E. Gallo, J. Bregman, K. Gultekin, H. W. Rix, S. Faber, S. Ellison, A. Pontzen, L. Sales, B. O'Shea, M. Donahue, and M. Voit. This work used the SIMBAD database at CDS, and the NASA/IPAC Extragalactic Database (NED), operated by the Jet Propulsion Laboratory, California Institute of Technology, under contract with NASA.

\onecolumngrid
\newpage

\begin{deluxetable}{lcccclc}
\tablecaption{Columns: (1) Galaxy name, (2) stellar mass derived from 2MASS $K_{\rm{s}}$ apparent magnitudes where we assume an error of 0.15 dex, (3) star formation rate derived from far-infrared measurements where values with an asterisk indicate upper limits and where we assume an error of 0.3 dex, (4)-(5) black hole mass and error, (6) black hole mass measurement method - either stellar dynamics, CO or gas dynamics, masers, or reverberation mapping, (7) reference for black hole measurement: 1 = \citet{soe2016}; 2 = \citet{v2016}.}
\tablenum{1}
\label{table1}

\tablehead{\colhead{(1)} & \colhead{(2)} & \colhead{(3)} & \colhead{(4)} & \colhead{(5)} & \colhead{(6)} & \colhead{(7)} \\
\colhead{Name} & \colhead{$M_{*}$} & \colhead{SFR} & \colhead{$M_{\rm{BH}}$} & \colhead{$M_{\rm{BH}}$ error} & \colhead{method} & \colhead{ref} \\ 
\colhead{} & \colhead{(log$_{10}$ M$_{\odot}$)} & \colhead{(log$_{10}$ M$_{\odot}$ yr$^{-1}$)} & \colhead{(log$_{10}$ M$_{\odot}$)} & \colhead{(log$_{10}$ M$_{\odot}$)} & \colhead{} & \colhead{} } 

\startdata
Centaurus A & 10.904 & 0.213 & 7.755 & 0.084 & star & 1 \\
Circinus & 10.200 & -0.010 & 6.057 & 0.105 & maser & 1 \\
IC1459 & 11.381 & -0.611 & 9.394 & 0.079 & star & 1 \\
IC4296 & 11.567 & -0.753 & 9.114 & 0.073 & gas & 1 \\
M31 & 10.731 & -0.519 & 8.155 & 0.161 & star & 1 \\
M66 & 10.840 & 0.536 & 6.929 & 0.048 & star & 1 \\
M81 & 10.764 & -0.356 & 7.813 & 0.129 & star, gas & 1 \\
M87 & 11.519 & -1.335 & 9.789 & 0.031 & star & 1 \\
NGC0307 & 10.772 &  -0.567$^{*}$ & 8.602 & 0.060 & star & 1 \\
NGC0524 & 11.086 & -0.559 & 8.938 & 0.053 & star & 1 \\
NGC0821 & 10.779 & -1.189 & 8.217 & 0.210 & star & 1 \\
NGC1023 & 10.756 &  -0.730$^{*}$ & 7.616 & 0.055 & star & 1 \\
NGC1068 & 11.271 & 1.304 & 6.924 & 0.245 & maser & 1 \\
NGC1194 & 10.806 & 0.194 & 7.850 & 0.051 & maser & 1 \\
NGC1316 & 11.594 & -0.187 & 8.176 & 0.254 & star & 1 \\
NGC1332 & 11.060 & -0.739 & 9.161 & 0.076 & star & 1 \\
NGC1398 & 11.375 & -0.067 & 8.033 & 0.083 & star & 1 \\
NGC1399 & 11.297 & -1.513 & 8.945 & 0.306 & star & 1 \\
NGC1407 & 11.495 & -1.110 & 9.653 & 0.079 & star & 1 \\
NGC1550 & 11.100 &  -0.588$^{*}$ & 9.568 & 0.067 & star & 1 \\
NGC2273 & 10.731 & 0.416 & 6.935 & 0.036 & maser & 1 \\
NGC2549 & 10.172 & -1.706 & 7.161 & 0.367 & star & 1 \\
NGC2787 & 10.022 & -1.750 & 7.610 & 0.088 & gas & 1 \\
NGC2960 & 10.925 & 0.265 & 7.033 & 0.049 & maser & 1 \\
NGC2974 & 11.354 & -0.791 & 8.230 & 0.091 & star & 1 \\
NGC3079 & 10.683 & 0.782 & 6.398 & 0.049 & maser & 1 \\
NGC3115 & 10.789 & -2.146 & 8.953 & 0.095 & star & 1 \\
NGC3227 & 10.882 & 0.398 & 7.322 & 0.232 & star & 1 \\
NGC3245 & 10.698 & -0.306 & 8.378 & 0.114 & gas & 1 \\
NGC3368 & 10.689 & -0.222 & 6.875 & 0.076 & star & 1 \\
NGC3393 & 10.943 & 0.439 & 7.196 & 0.330 & maser & 1 \\
NGC3414 & 10.797 & -1.047 & 8.400 & 0.071 & star & 1 \\
NGC3585 & 11.126 & -1.391 & 8.517 & 0.127 & star & 1 \\
NGC3607 & 11.097 & -0.319 & 8.137 & 0.157 & star & 1 \\
NGC3842 & 11.577 & 0.382 & 9.959 & 0.139 & star & 1 \\
NGC3923 & 11.234 &  -1.766$^{*}$ & 9.449 & 0.115 & star & 1 \\
NGC3998 & 10.548 & -1.210 & 8.927 & 0.052 & star & 1 \\
NGC4151 & 10.837 & 0.051 & 7.813 & 0.076 & star & 1 \\
NGC4258 & 10.721 & -0.080 & 7.577 & 0.030 & maser & 1 \\
NGC4261 & 11.427 & -1.359 & 8.723 & 0.097 & gas & 1 \\
NGC4291 & 10.664 &  -1.193$^{*}$ & 8.990 & 0.155 & star & 1 \\
NGC4472 & 11.663 &  -1.337$^{*}$ & 9.398 & 0.037 & star & 1 \\
NGC4594 & 11.322 & -0.412 & 8.823 & 0.045 & star & 1 \\
NGC4697 & 10.832 & -1.349 & 8.305 & 0.112 & star & 1 \\
NGC4699 & 11.140 & 0.125 & 8.246 & 0.052 & star & 1 \\
NGC4736 & 10.539 & -0.061 & 6.831 & 0.123 & star & 1 \\
NGC4826 & 10.774 & 0.009 & 6.193 & 0.131 & star & 1 \\
NGC4889 & 11.836 &  0.034$^{*}$ & 10.320 & 0.437 & star & 1 \\
NGC5018 & 11.319 & -0.080 & 8.021 & 0.078 & star & 1 \\
NGC5077 & 11.070 & -0.976 & 8.932 & 0.268 & gas & 1 \\
NGC5328 & 11.410 &  -0.390$^{*}$ & 9.672 & 0.158 & star & 1 \\
NGC5419 & 11.686 & -0.767 & 9.860 & 0.144 & star & 1 \\
NGC5846 & 11.204 & -1.453 & 9.041 & 0.058 & star & 1 \\
NGC6086 & 11.475 & 0.310$^{*}$ & 9.573 & 0.167 & star & 1 \\
NGC6251 & 11.641 & 0.707$^{*}$ & 8.788 & 0.155 & gas & 1 \\
NGC7052 & 11.450 & 0.176 & 8.598 & 0.230 & gas & 1 \\
NGC7457 & 10.107 & -1.869 & 6.954 & 0.302 & star & 1 \\
NGC7582 & 10.953 & 1.066 & 7.741 & 0.205 & gas & 1 \\
NGC7619 & 11.395 & -0.353 & 9.398 & 0.108 & star & 1 \\
NGC7768 & 11.576 & 0.151$^{*}$ & 9.127 & 0.181 & star & 1 \\
UGC3789 & 10.775 & 0.351 & 6.985 & 0.085 & maser & 1 \\
3C120 & 11.448 & 1.155 & 7.730 & 0.150 & RM & 2 \\
Ark120 & 11.555 & 0.800 & 8.050 & 0.170 & RM & 2 \\
IC1481 & 10.843 & 0.689 & 7.150 & 0.130 & maser & 2 \\
Mrk110 & 10.822 & 0.257 & 7.280 & 0.210 & RM & 2 \\
Mrk279 & 11.384 & 1.035 & 7.400 & 0.230 & RM & 2 \\
Mrk290 & 10.687 & -0.029 & 7.260 & 0.170 & RM & 2 \\
Mrk335 & 11.251 & 0.403 & 7.210 & 0.160 & RM & 2 \\
Mrk509 & 11.525 & 1.108 & 8.030 & 0.150 & RM & 2 \\
Mrk590 & 11.478 & 0.610 & 7.550 & 0.180 & RM & 2 \\
Mrk79 & 11.230 & 0.820 & 7.580 & 0.230 & RM & 2 \\
Mrk817 & 11.304 & 1.216 & 7.570 & 0.180 & RM & 2 \\
NGC1600 & 12.175 & -0.077 & 10.230 & 0.040 & star & 2 \\
NGC3516 & 10.942 & 0.066 & 7.370 & 0.160 & RM & 2 \\
NGC3783 & 10.981 & 0.434 & 7.360 & 0.190 & RM & 2 \\
NGC4253 & 10.737 & 0.734 & 6.800 & 0.170 & RM & 2 \\
NGC4593 & 11.164 & 0.381 & 6.860 & 0.210 & RM & 2 \\
NGC5273 & 10.099 & -0.961 & 6.610 & 0.270 & RM & 2 \\
NGC5548 & 11.165 & 0.445 & 7.700 & 0.130 & RM & 2 \\
NGC5765b & 11.209 & 1.434 & 7.660 & 0.030 & maser & 2 \\
NGC6814 & 10.843 & 0.286 & 7.020 & 0.170 & RM & 2 \\
NGC7469 & 11.011 & 1.434 & 6.940 & 0.160 & RM & 2 \\
NGC1097 & 11.009 & 0.766 & 8.140 & 0.090 & CO & 2 \\
NGC1275 & 11.646 & 1.109 & 8.980 & 0.200 & gas & 2 \\
NGC3665 & 11.194 & 0.246 & 8.760 & 0.090 & CO & 2 \\
NGC3706 & 11.361 & -1.002 & 9.770 & 0.060 & star & 2 \\
NGC4303 & 10.955 & 0.674 & 6.510 & 0.740 & gas & 2 \\
NGC4742 & 10.226 & -1.175 & 7.100 & 0.150 & star & 2 \\
NGC5495 & 11.392 & 0.963 & 7.080 & 0.300 & maser & 2 \\
NGC7332 & 10.656 & -1.279 & 7.080 & 0.180 & star & 2 \\
\enddata

\end{deluxetable}


\begin{deluxetable}{lcccccccccc}
\tabletypesize{\footnotesize}
\tablecaption{Columns: (1) Galaxy name; (2)-(3) distances and errors taken from the same references as the black hole masses, see Column 7 in Table~\ref{table1}; (4) extinction-corrected 2MASS `total' $K_{\rm{s}}$ apparent magnitudes from \citet{hmm2012} unless the value has an asterisk in which case these are taken from the 2MASS LGA Catalog \citep{jcc2003}; (5)-(8) IRAS 12$\mu$, 25$\mu$, 60$\mu$, 100$\mu$ flux measurements; (9) MIPS 70$\mu$ flux measurement; (10) IRAS measurement reference: 1 = \citet{rls1988}, 2 = corrections to \citet{kgk1989} in NED by Knapp (1994), 3 = \citet{m1990}, 4 = \citet{ssm2004}, 5 = \citet{gbi1992}, 6 = \citet{sh2009}; (11) MIPS measurement reference: 1 = \citet{tbm2009}, 2 = \citet{dcj2009}, 3 = \citet{ttt2010}, 4 = \citet{erg2008}, 5 = \citet{sbw2011}, 6 = \citet{lbb2010}. If there is a reference but no flux measurement then this is taken to be a non-detection.
\newline
$^{1}$The SFRs for NGC 1023 and NGC6251 are calculated based on 12$\mu$m and 25$\mu$m detections and therefore may be contaminated by an AGN. We therefore use these detections to obtain upper limits on these galaxies' SFRs.
\newline
$^{\dagger}$NGC7768 has an IRAS 25$\mu$m detection yet is likely contaminated by a nearby star, therefore we omit this value since it is clearly an elliptical BCG.}

\tablenum{2}
\label{table2}

\tablehead{\colhead{(1)} & \colhead{(2)} & \colhead{(3)} & \colhead{(4)} & \colhead{(5)} & \colhead{(6)} & \colhead{(7)} & \colhead{(8)} & \colhead{(9)} & \colhead{(10)} & \colhead{(11)} \\
\colhead{Name} & \colhead{Distance} & \colhead{Distance error} & \colhead{$K_{\rm{s}}$} & \colhead{IRAS 12$\mu$} & \colhead{IRAS 25$\mu$} & \colhead{IRAS 60$\mu$} & \colhead{IRAS 100$\mu$} & \colhead{MIPS 70$\mu$} & \colhead{IRAS ref} & \colhead{MIPS ref} \\ 
\colhead{} & \colhead{(Mpc)} & \colhead{(Mpc)} & \colhead{} & \colhead{(Jy)} & \colhead{(Jy)} & \colhead{(Jy)} & \colhead{(Jy)} & \colhead{(Jy)} & \colhead{} & \colhead{} } 

\startdata
Centaurus A & 3.620 & 0.200 & 3.490 & 23.030 & 30.740 & 217.570 & 501.200 & - & 1 & - \\
Circinus & 2.820 & 0.470 & 4.710 & 18.800 & 68.440 & 248.700 & 315.850 & - & 5 & - \\
IC1459 & 28.920 & 3.739 & 6.810 & 0.170 & 0.320 & 0.510 & 1.180 & 0.542 & 2 & 1 \\
IC4296 & 49.200 & 3.628 & 7.500 & - & - & 0.140 & 0.260 & 0.118 & 2 & 1 \\
M31 & 0.774 & 0.032 & 0.573 & 163.000 & 108.000 & 536.000 & 2928.400 & 1200.000 & 1 & 3 \\
M66 & 10.050 & 1.092 & 5.869$^{*}$ & 4.170 & 7.720 & 56.310 & 144.960 & 91.900 & 1 & 2 \\
M81 & 3.604 & 0.133 & 3.831 & 5.860 & 5.420 & 44.700 & 174.000 & 85.300 & 1 & 2 \\
M87 & 16.680 & 0.615 & 5.270 & 0.440 & 0.187 & 0.390 & 0.410 & 0.483 & 2 & 1 \\
NGC0307 & 52.800 & 5.736 & 9.641 & - & - & - & - & - & 2 & - \\
NGC0524 & 24.220 & 2.234 & 7.163 & 0.240 & - & 0.760 & 2.050 & - & 2 & - \\
NGC0821 & 23.440 & 1.837 & 7.861 & - & - & - & 0.500 & - & 2 & 1 \\
NGC1023$^{1}$ & 10.810 & 0.797 & 6.238 & 0.240 & - & - & - & - & 2 & 1 \\
NGC1068 & 15.900 & 9.411 & 5.788 & 39.700 & 85.040 & 176.200 & 224.000 & - & 3 & - \\
NGC1194 & 57.980 & 6.299 & 9.758 & 0.266 & 0.512 & 0.770 & - & - & 3 & - \\
NGC1316 & 18.600 & 0.600 & 5.320 & 0.330 & 0.290 & 3.070 & 8.110 & 5.440 & 2 & 1 \\
NGC1332 & 22.300 & 1.851 & 7.050 & 0.100 & 0.110 & 0.510 & 1.810 & - & 2 & - \\
NGC1398 & 24.770 & 4.125 & 6.49$^{*}$ & 0.143 & 0.116 & 1.141 & 8.963 & - & 3 & - \\
NGC1399 & 20.850 & 0.672 & 6.310 & 0.100 & - & - & 0.300 & 0.016 & 2 & 1 \\
NGC1407 & 28.050 & 3.367 & 6.460 & 0.120 & - & 0.140 & 0.480 & - & 2 & 1 \\
NGC1550 & 51.570 & 5.603 & 8.770 & - & - & - & - & - & 2 & - \\
NGC2273 & 29.500 & 1.903 & 8.480 & 0.400 & 1.360 & 6.020 & 10.000 & - & 3 & - \\
NGC2549 & 12.700 & 1.642 & 8.046 & - & - & 0.260 & 0.370 & - & 2 & - \\
NGC2787 & 7.450 & 1.241 & 7.263 & 0.080 & 0.100 & 0.600 & 1.180 & 1.017 & 2 & 1 \\
NGC2960 & 67.100 & 7.120 & 9.780 & - & - & 0.708 & 1.657 & - & 3 & - \\
NGC2974 & 21.500 & 2.381 & 6.236 & - & - & 0.420 & 1.900 & 0.682 & 2 & 1 \\
NGC3079 & 15.900 & 1.246 & 7.258 & 1.523 & 2.272 & 44.500 & 89.200 & 63.700 & 3 & 4 \\
NGC3115 & 9.540 & 0.396 & 5.883 & 0.360 & 0.110 & 0.130 & - & 0.052 & 2 & 1 \\
NGC3227 & 23.750 & 2.630 & 7.631 & 0.667 & 1.764 & 7.825 & 17.590 & - & 3 & - \\
NGC3245 & 21.380 & 1.972 & 7.862 & 0.160 & 0.230 & 2.030 & 3.970 & - & 2 & - \\
NGC3368 & 10.400 & 0.959 & 6.320 & 0.535 & 0.544 & 8.261 & 25.930 & 14.500 & 3 & 2 \\
NGC3393 & 49.200 & 8.194 & 9.059 & 0.131 & 0.753 & 2.251 & 3.873 & - & 3 & - \\
NGC3414 & 25.200 & 2.738 & 7.972 & 0.080 & - & 0.250 & 0.560 & - & 2 & - \\
NGC3585 & 20.510 & 1.702 & 6.703 & 0.120 & 0.210 & 0.160 & - & 0.080 & 2 & 1 \\
NGC3607 & 22.650 & 1.775 & 6.990 & - & - & - & - & 1.761 & - & 1 \\
NGC3842 & 92.200 & 10.638 & 8.840 & 0.090 & - & 0.360 & 1.490 & - & 2 & - \\
NGC3923 & 20.880 & 2.700 & 6.471$^{*}$ & 0.130 & - & - & - & 0.024 & 2 & 1 \\
NGC3998 & 14.300 & 1.253 & 7.365 & 0.140 & 0.130 & 0.550 & 1.150 & - & 2 & - \\
NGC4151 & 20.000 & 2.772 & 7.371 & 1.970 & 4.830 & 6.320 & 7.640 & - & 3 & - \\
NGC4258 & 7.270 & 0.503 & 5.464 & 2.250 & 2.810 & 21.600 & 78.390 & 40.700 & 1 & 2 \\
NGC4261 & 32.360 & 2.835 & 6.940 & 0.180 & 0.090 & 0.080 & 0.150 & 0.127 & 2 & 1 \\
NGC4291 & 26.580 & 3.931 & 8.420 & - & - & - & - & - & 2 & 1 \\
NGC4472 & 17.140 & 0.592 & 4.970 & 0.200 & - & - & - & 0.061 & 2 & 1 \\
NGC4594 & 9.870 & 0.819 & 4.625 & 0.740 & 0.500 & 4.260 & 22.900 & 7.310 & 1 & 2 \\
NGC4697 & 12.540 & 0.404 & 6.370 & 0.290 & - & 0.460 & 1.240 & 0.618 & 2 & 1 \\
NGC4699 & 18.900 & 2.053 & 6.489 & 0.383 & 0.340 & 4.979 & 19.020 & - & 3 & - \\
NGC4736 & 5.000 & 0.786 & 5.106 & 4.770 & 6.830 & 62.410 & 135.340 & 101.000 & 1 & 2 \\
NGC4826 & 7.270 & 1.177 & 5.330 & 1.710 & 2.000 & 33.860 & 77.380 & 52.900 & 1 & 2 \\
NGC4889 & 102.000 & 5.169 & 8.410 & - & - & - & - & - & 2 & - \\
NGC5018 & 40.550 & 4.867 & 7.700 & 0.200 & - & 0.950 & 1.860 & 1.174 & 2 & 1 \\
NGC5077 & 38.700 & 8.442 & 8.220 & - & - & - & - & 0.133 & - & 1 \\
NGC5328 & 64.100 & 6.964 & 8.467 & - & - & - & - & - & 2 & - \\
NGC5419 & 56.200 & 6.106 & 7.492$^{*}$ & - & - & - & 0.230 & - & 2 & 1 \\
NGC5846 & 24.900 & 2.297 & 6.929 & - & - & - & - & 0.107 & 2 & 1 \\
NGC6086 & 138.000 & 11.452 & 9.970 & - & - & - & - & - & 2 & - \\
NGC6251$^{1}$ & 108.400 & 8.996 & 9.030 & - & 0.100 & - & - & - & 2 & - \\
NGC7052 & 70.400 & 8.449 & 8.570 & 0.050 & - & 0.450 & 1.420 & - & 2 & - \\
NGC7457 & 12.530 & 1.214 & 8.179 & - & - & 0.110 & 0.450 & - & 2 & 1 \\
NGC7582 & 22.300 & 9.845 & 7.316 & 1.620 & 6.436 & 49.100 & 72.920 & - & 3 & - \\
NGC7619 & 51.520 & 7.380 & 8.030 & - & - & - & 0.710 & - & 2 & 1 \\
NGC7768 & 116.000 & 27.495 & 9.340 & - & - & - & - & - & 2$^{\dagger}$ & - \\
UGC3789 & 49.900 & 5.421 & 9.510 & 0.111 & 0.329 & 1.669 & 3.377 & - & 3 & - \\
3C120 & 141.400 & 14.100 & 10.089 & 0.286 & 0.635 & 1.283 & 2.786 & - & 3 & - \\
Ark120 & 140.100 & 14.000 & 9.803 & 0.319 & 0.410 & 0.643 & 1.084 & - & 3 & - \\
IC1481 & 89.900 & 9.000 & 10.62$^{*}$ & - & 0.275 & 1.410 & 1.510 & - & 3 & - \\
Mrk110 & 151.100 & 15.100 & 11.8$^{*}$ & - & - & 0.131 & - & - & 6 & - \\
Mrk279 & 130.400 & 13.000 & 10.073 & - & - & 1.255 & 2.200 & - & 3 & - \\
Mrk290 & 126.700 & - & 11.754$^{*}$ & - & - & 0.140 & 0.136 & 0.172 & 6 & 5 \\
Mrk335 & 110.500 & - & 10.047 & 0.302 & 0.378 & 0.343 & - & - & 3 & - \\
Mrk509 & 147.300 & 14.700 & 9.985 & 0.316 & 0.702 & 1.364 & 1.521 & 1.440 & 3 & 5 \\
Mrk590 & 113.000 & 11.300 & 9.527 & 0.192 & 0.221 & 0.489 & 1.457 & - & 3 & - \\
Mrk79 & 95.000 & 9.500 & 9.772 & 0.306 & 0.763 & 1.503 & 2.363 & - & 3 & - \\
Mrk817 & 134.700 & 13.500 & 10.344 & 0.336 & 1.175 & 2.118 & 2.268 & - & 3 & - \\
NGC1600 & 126.300 & 11.600 & 8.026 & - & - & 0.100 & 0.190 & - & 2 & - \\
NGC3516 & 37.900 & 3.800 & 8.497 & 0.410 & 1.010 & 1.850 & 2.130 & - & 2 & - \\
NGC3783 & 41.700 & 4.200 & 8.606 & 0.840 & 2.492 & 3.257 & 4.899 & - & 3 & - \\
NGC4253 & 55.400 & 5.500 & 9.832 & 0.386 & 1.300 & 4.030 & 4.660 & - & 3 & - \\
NGC4593 & 38.500 & 3.900 & 7.976 & 0.344 & 0.809 & 3.052 & 5.947 & - & 3 & - \\
NGC5273 & 15.500 & 1.500 & 8.661 & 0.120 & 0.290 & 0.900 & 1.560 & 0.657 & 2 & 1 \\
NGC5548 & 73.600 & 7.400 & 9.380 & 0.401 & 0.769 & 1.073 & 1.614 & - & 3 & - \\
NGC5765b & 126.300 & 12.600 & 10.442 & 0.291 & 0.752 & 3.367 & 5.830 & - & 3 & - \\
NGC6814 & 22.300 & 2.200 & 7.591 & - & 0.599 & 5.517 & 18.880 & - & 3 & - \\
NGC7469 & 47.700 & 8.100 & 8.823 & 1.348 & 5.789 & 25.870 & 34.900 & - & 4 & - \\
NGC1097 & 14.500 & - & 6.243 & 2.880 & 7.700 & 46.730 & 116.340 & 59.840 & 1 & 2 \\
NGC1275 & 70.000 & - & 8.068 & 0.950 & 2.830 & 5.760 & 7.500 & 3.990 & 2 & 6 \\
NGC3665 & 34.700 & - & 7.675 & 0.110 & 0.210 & 1.910 & 7.530 & - & 2 & - \\
NGC3706 & 46.000 & - & 7.869 & 0.080 & - & 0.070 & 0.220 & - & 2 & 1 \\
NGC4303 & 17.900 & - & 6.835 & 1.064 & 1.401 & 23.640 & 64.650 & - & 3 & - \\
NGC4742 & 15.700 & - & 8.371 & - & - & 0.450 & 1.150 & - & 2 & - \\
NGC5495 & 103.000 & - & 9.543 & 0.112 & - & 1.487 & 3.534 & - & 3 & - \\
NGC7332 & 21.700 & - & 7.999 & 0.120 & - & 0.210 & 0.410 & - & 2 & - \\
\enddata

\end{deluxetable}

\end{document}